\documentclass{aa}
\usepackage{psfig}

\def\solar {\ifmmode_{\mathord\odot} \else $_{\mathord\odot}$\fi} 
\def\Msol {\ifmmode {\,{\it M}\solar} \else $\,M$\solar\fi}        
\def\deg{\ifmmode^\circ\else$^\circ$\fi}
\begin{document}

\thesaurus{
  06         
  (
  08.02.3;  
  08.02.4;  
  08.02.6;  
  08.12.2)  
  } 
\title{
  Accurate masses of very low mass stars: \\
I Gl~570BC (0.6~{\Msol}+0.4~{\Msol}) 
 \thanks{
    Partly based on observations made at Observatoire de Haute-Provence,
    operated by the Centre National de la Recherche Scientifique de France,
    on observations made at Canada-France-Hawaii Telescope, operated by 
    the National Research Council of Canada, the Centre National de la 
    Recherche Scientifique de France and the University of Hawaii, at the
    3.6m telescope of the European Southern Observatory at La Silla (Chile),
    and the Kitt Peak Mayall 4m telescope operated by NOAO.
    }
  } 

\author{
  Thierry Forveille \inst{1}, 
  Jean-Luc Beuzit \inst{2,1,3}, 
  Xavier Delfosse \inst{1,3,4}, 
  Damien Segransan \inst{1}, 
  Francoise Beck \inst{1}, 
  Michel Mayor \inst{3}, 
  Christian Perrier \inst{1}, 
  Andrei Tokovinin \inst{5},
  St\'ephane Udry \inst{3}
  }
\authorrunning{Forveille et al.}
\titlerunning{Accurate masses of two early M dwarfs}
\offprints{Thierry Forveille, e-mail: Thierry.Forveille@obs.ujf-grenoble.fr}

\institute{
  Observatoire de Grenoble,
  414 rue de la Piscine,
  Domaine Universitaire de S$^{\mathrm t}$ Martin d'H\`eres,
  F-38041 Grenoble,
  France
  \and
  Canada-France-Hawaii Telescope Corporation, 
  P.O. Box 1597,
  Kamuela, HI 96743, 
  U.S.A.
  \and
  Observatoire de Gen\`eve
  CH-1290 Sauverny,
  Switzerland
  \and
  Instituto de Astrofisica de Canarias, 
  E-38200 La Laguna (Tenerife),
  Spain
  \and
  Sternberg Astronomical Institute, 
  Universitetsky prosp. 13, 
  119899 Moscow, 
  Russia 
  }

\date{Received ; accepted }

\maketitle

\begin{abstract}

We present very accurate individual masses (1.2\% relative accuracy)
for the two components of Gl~570BC, an interferometric and double-lined 
spectroscopic binary system. They were obtained from new high 
accuracy radial velocity and angular separation measurements,
analysed together with previously published
measurements. From those data we determine a much improved orbit
through a simultaneous least square fit to the radial velocity,
visual, and parallax information. The derived masses and absolute
magnitudes generally validate the theoretical and empirical
mass-luminosity relations around 0.5~{\Msol}, but point towards some
low level discrepancies at the 0.1 to 0.15 magnitude level. Forthcoming
results of this observing program will extend the comparison to much
lower masses with similar accuracy.

\keywords{stars: binaries - 
  stars: low mass, brown dwarfs -
  stars: individual: Gl 570 -
  techniques: radial velocity - 
  techniques: adaptive optics }

\end{abstract}

\section{Introduction}
Fundamental determination of stellar masses from binary orbits is a
most classical astrophysical discipline, last comprehensively reviewed
by Andersen (1991; 1998). Besides defining a mean mass-luminosity
relation, used at many places in astronomy to approximately convert the
observable stellar light to the underlying mass (for instance to
derive an initial mass function), accurate stellar masses in multiple
systems provide what is perhaps the most demanding and fundamental
test of stellar evolution theory (e.g. Andersen 1991). Mass, the
basic input of evolutionary models, is directly measured and the
models must reproduce the effective temperatures and radii (or
luminosities) of both components, for a single age and a single
chemical composition. Given the strong dependence of all stellar
parameters on mass, this discriminating diagnostic however only shows
its power for relative mass errors $\leq$1-2\%. In practice this has
up to now restricted its use to double-lined detached eclipsing
binaries. 
These systems are however relatively rare: only 44 have yielded masses
accurate enough to be included in Andersen's (1991) critical
compilation, and few have appeared in the literature since
then. Tidally induced rotational mixing may in addition affect the
evolution of the short period eclipsing systems, perhaps sufficiently
that they are not completely representative of isolated stars. More
seriously however, detached eclipsing systems poorly fill some
interesting areas in the HR diagram.

The lower main sequence is one major region with very few known
eclipsing systems, as a result of the strong observational and
intrinsic biases against observing eclipses in faint and physically
small stars. Only three well detached
eclipsing binaries are known with significantly subsolar component
masses: YY Gem (M0Ve, 0.6+0.6 \Msol; Bopp 1974; Leung \& Schneider
1978), CM Dra (M4Ve, 0.2+0.2 \Msol; Lacy 1977, Metcalfe et al. 1996),
and the recently identified GJ~2069A (M3.5Ve, 0.4+0.4 \Msol; Delfosse
et al. 1999a).  Detailed observational checks of evolutionary models
(e.g. Paczynski \& Sienkiewicz 1984; Chabrier \& Baraffe 1995) have
therefore heavily relied on the first two of these systems, even
though they are in some respects non-ideal for this purpose: both
binaries have two nearly equal mass components, so that the strength
of the differential comparison of two stars with different masses but
otherwise equal parameters is largely lost. Also, all three are
chromospherically very active, due to tidal synchronisation of their
rotation with the short orbital period. As a consequence, they may
have untypical colours for their bolometric luminosity.

Angularly resolved spectroscopic binaries provide stellar masses in
parts of the HR diagram where eclipsing systems are missing, though to
date these systems have not matched the $\sim$1\% accuracy which can
be obtained in detached eclipsing systems. For M dwarfs in particular,
the best representation to date of the empirical M-L relation (Henry
\& Mc~Carthy 1993, hereafter HMcC; Henry, Franz, Wasserman et al.,
1999) still has to rely in part on some fairly noisy mass
determinations.  For several years (Perrier et al. 1992) we have
therefore been following up with high angular resolution some low mass
spectroscopic systems found with the CORAVEL or ELODIE radial velocity
spectrographs. This follow-up initially used one-dimensional (1D) IR
speckle, then two-dimensional (2D) IR speckle, and now uses adaptive
optics imaging. As a progress report on this program and as an
illustration of our methods, we present here much improved parameters
for the double-lined interferometric binary Gl~570BC. The $\sim$1\%
accuracy for the 2 masses improves by an order of magnitude on our
earlier measurements (Mariotti et al. 1990) of this system, and is
getting close to what is obtainable for eclipsing systems.

The Gl~570 system comprises the {\it V}=5.7 K4V star Gl~570A (also HR~5568,
HD~131977, HIP~73184, FK5 1391), and at a projected distance of 25$''$
the close Gl~570BC pair (also HD 131976, HIP~73182) which is the
subject of the present paper. As discussed below, the orbital period of
the close pair is fairly short, only 10~months. Thanks to its small
distance of only 6~pc it is nonetheless usually well resolved by
the diffraction limit of 4m-class telescopes. The separation within
Gl~570BC on the other hand always remains less than 0.2$''$, so that all
seeing-limited mesurements refer to integrated properties of the close
pair. Its integrated magnitude is {\it V}=8.09, and its joint spectral type
is M1V (Henry et al. 1994; Reid et al. 1995). The three components have
common parallax, proper motion, and radial velocity. They are
therefore gravitationnally bound, with the projected separation of the
wide pair (about 120~AU at the distance of the system) corresponding
to P~$\sim$~500~years. Formation of the system could either result
from the fragmentation of a single gas clump or have involved some
capture(s). The latter process however only remains efficient at the
high densities which characterize very rich star forming clusters,
whose lifetimes are very much shorter than the hydrogen burning
timescale in a K4 dwarf. For all practical purposes, the three stars
can thus in both cases be considered as coeval and as formed from the
same gas.
 
\section{Observations and results}
\subsection{Angular separation measurements}
A variety of astrometric observations of different kinds and of
different quality have been collected for Gliese~570BC over the
years. The close pair was first resolved by Mariotti et al.  (1990) who
obtained infrared 1D speckle observations on three occasions and
derived a first visual orbit. We refer to their paper for the
description of the instrument and the observing and data reduction
procedures.  HMcC later measured Gl~570BC once with a 2D infrared speckle
instrument. For readers' convenience, we list these published measurements
in Table~\ref{tab_oa} together with our new 2D speckle and adaptive optics 
(hereafter AO) observations.

Two measurements were obtained in February 1991 and April 1991 using
the speckle mode of the 2D infrared imagers then installed
respectively at the KPNO 4.2m telescope and the CFHT 3.6m telescope.
Each imager was designed to permit acquisition of exposures short
enough to substantially freeze the seeing under standard atmospheric
conditions ($t_{exp}\approx$ 50 to 100~ms) in bands {\it H} and {\it K},
independently of the overhead due to read-out time and data transfer
time.  Several sequences of a few hundred such short exposures were
obtained, alternating every few minutes between the source and a
nearby unresolved star (usually Gl~570A) used as a point spread function (PSF)
calibrator. The whole observation took one hour or less.

In principle, this observing procedure allows an almost simultaneous
PSF calibration, and consequently estimates the visibility modulus
with 1 to 5\% accuracy. We used a software package written
specifically for this type of data reduction by E. Tessier (Tessier et
al. 1994). It produces an unbiased visibility estimator for the source
and then extracts the binary parameters and their estimated variance
from this visibility.
The actual detector scale and position angle (P.A.) origin for those
observations were calibrated from observations of the astrometric
binary $\zeta$~Aqr (Heintz 1989).

\begin{table}
  \center
  \tabcolsep 0.1cm
  \caption{Angular separation measurements}
  \begin{tabular}{crcc}
Date    & Pos. Angle & Separation & Ref.\\
        & degree     & arc second & \\ 
    46961 & 0 & -0.060$\pm$0.010 & 1\\
    46961 & 90 & -0.090$\pm$0.010 & 1\\
    46961 & 45 & -0.110$\pm$0.010 & 1\\
    46961 & 135 & -0.020$\pm$0.040 & 1\\
    47071 & 0 & -0.154$\pm$0.025 & 1\\
    47071 & 90 & -0.097$\pm$0.025 & 1\\
    47274 & 0 & -0.075$\pm$0.020 & 1\\
    47274 & 45 & -0.114$\pm$0.010 & 1\\
    47274 & 90 & -0.094$\pm$0.020 & 1\\
    47343 & 0 & -0.130$\pm$0.010 & 1\\
    47343 & 45 & -0.173$\pm$0.010 & 1\\
    47343 & 90 & -0.114$\pm$0.020 & 1 \vspace{5mm} \\

Date    & $\rho$ (arcsec) & $\theta$ (degree) & Ref. \\
47934 & 0.153$\pm$0.006 & 220.0$\pm$2.0 & 2 \\
48312 & 0.194$\pm$0.010 & 210.2$\pm$3.0 & 3 \\
48372 & 0.146$\pm$0.020 & 203.0$\pm$5.0 & 4 \\
49470 & 0.159$\pm$0.004 & 222.4$\pm$1.0 & 5 \\
49798 & 0.168$\pm$0.004 & 214.7$\pm$2.0 & 6 \\
50499 & 0.193$\pm$0.002 & 206.5$\pm$0.2 & 7\\
50810 & 0.181$\pm$0.004 & 205.9$\pm$0.8 & 7\\
51017 & 0.142$\pm$0.005 & 221.0$\pm$2.0 & 7 \\
51017 & 0.147$\pm$0.006 & 223.0$\pm$2.0 & 7\\

  \end{tabular}
\raggedright
\noindent
\vskip 2mm
Notes: Observation dates are listed as offsets relative to Julian Day
2400000. 
\noindent
\vskip 2mm
References: 1) Mariotti et al., 1990; 2) Henry \& Mc Carthy 1993; 3)
This paper, Kitt Peak, 2D speckle imager; 4) This paper, ESO 3.6m,
COME-ON adaptive optics system; 5) CFHT, CIRCUS 2D speckle imager; 6)
ESO 3.6m, ADONIS adaptive optics system; 7) This paper, CFHT, PUE'O
adaptive optics system.
  \label{tab_oa}
\end{table}

Most of the new measurements were obtained at the 3.6m
Canada-France-Hawaii Telescope (CFHT) on top Mauna Kea,
using the CFHT Adaptive Optics
Bonnette (Arsenault et al. 1994, Rigaut et al. 1998) and two different
infrared cameras (Nadeau et al. 1994, Doyon et al. 1998). Delfosse et
al. (1999b) provide a detailed description of the observing procedure,
\linebreak which we therefore don't repeat here.  Two observations were also
obtained with the ESO 3.6m telescope (La Silla, Chile) AO system
COME-ON+ (Rousset \& Beuzit 1999), then ADONIS (Beuzit et al. 1999),
equipped with the SHARP-II infrared camera, using a similar observing
procedure.  This COME-ON+ observation has been published in Mariotti
et al (1991).  For recent CFHT measurements the corrected point spread
function obtained from the AO system was synthesized from simultaneous
recordings of the wavefront sensor measurements and deformable mirror
commands, as described by V\'eran et al. (1997). For pre-1997
measurements it was instead obtained from observations of a reference
single star of similar R-band magnitude.  Astrometric calibration
fields such as the central region of the Trapezium Cluster in the
Orion Nebula (McCaughrean \&  Stauffer 1994), were observed to
accurately determine the actual detector plate scale and position
angle (P.A.) origin. Uncertainties on these parameters do not
apreciably contribute to the overall separation error for the small
separations in the Gl~570BC system.
 
The separation, position angle and magnitude difference between the
two stars were determined using {\it uv} plane model fitting in the
GILDAS (Grenoble Image and Line Data Analysis System) software, as
well as with the
deconvolution algorithm described by V\'eran et al. (1999),
coded within IDL. With approximate initial values of the positions of
the two components along with the PSF reference image, the fitting
procedures gave as output the flux and pixel coordinates of the
primary and secondary.  Application of the astrometric calibrations
then yields the desired parameters.

\subsection{Radial velocity measurements}
\setcounter{table}{2}
%

All radial velocity measurements are listed in Table~2 
(available in electronic form only).  Most of them were obtained with
the two CORAVEL radial velocity scanners (Baranne et al. 1979) on
the Swiss 1m telescope at Observatoire de Haute Pro{\-}vence
(France) and on the Danish 1.54m telescope at La Silla (Chile). The
earlier data were previously published by Duquennoy \& Mayor (1988)
and the system has been regularly observed since then.  Gl~570BC is
relatively bright for the CORAVEL instruments (V=8.09, Leggett 1992),
but both stars are poor matches to the fixed K0III correlation
mask. They therefore only produce shallow correlation dips (of which
examples were displayed in Duquennoy \& Mayor (1988)) and their
velocities are measured with typical precisions of respectively 0.7
and 4~km/s, instead of the usual CORAVEL precision of 0.3~km/s. The
measurements of the M3V secondary in particular were at the limit of
the CORAVEL capabilities. During the orbit adjustment discussed below
they were found to have sizeable systematic errors at phases where the
two profiles are even slightly blended.  This didn't measurably affect
the derived orbital elements since these noisier measurements carried
essentially no weight anyway, but they were nonetheless ignored in the
final solution. For clarity, they are also not plotted in Figure~1.

Over the last three years, we have obtained considerably more accurate
measurements with the ELODIE spectrograph (Baranne et al. 1996) on the
1.93m telescope of Observatoire de Haute Pro{\-}vence and the CORALIE 
(Queloz et
al. in preparation) spectrograph on the Swiss 1.2m Euler telescope at
La Silla (Chile).  These echelle spectra were analysed by numerical
cross-correlation with an M4V one-bit (i.e. 0/1) mask, as described by
Delfosse et al. (1999b). Radial velocities were initially determined
by adjusting double gaussians to the correlation profiles. Those
however had systematic phase-dependent residuals during the orbital
adjustment, which were particularly large (400~m/s rms) for the
fainter secundary star. Upon analysis, they were found to originate
from the low level wiggles in the correlation profile of the primary:
while the core of this profile is well described by a gaussian
function, its baseline doesn't drop to zero as a gaussian would, but
instead keeps oscillating at the $\pm$0.2\% level, about one tenth of
the depth (2\%) of the secondary star's dip. Depending on its position
on this oscillating background, the measured velocity of the fainter
star could therefore be incorrect by up to 1~km/s. Similar errors of
course also affected the measured velocities of the primary star, but
they were smaller by the square of the relative depths of the two
correlation dips, or a factor of 16. Thanks in part to the excellent
stability of the spectrograph, the shape of the correlation profile
for a given star is very stable. We could therefore obtain an
excellent estimate of the wings of the intrinsic correlation profile
of each star, by averaging all profiles of the system, after aligning
them at the measured velocity of the star and blanking all pixels
within two profile widths of the velocity of the other star.
The residuals of a gaussian adjustment to these average profiles are
then subtracted from all correlation profiles, at the measured
velocity of each star. This decreases the fluctuation level in the
profile baseline to a level of $\pm$0.05\%.  The radial velocities
measured by a double gaussian fit to these corrected profiles have
typical accuracies of 40~m/s for the primary and 100~m/s for the
fainter secondary. These residuals are twice larger than would be
measured for single-lined spectra with equivalent S/N ratios and
correlation dip parameters, but show no systematic phase
dependency. They should thus cause no systematic errors on measured
parameters. 

\section{Orbit adjustment}
\subsection{The adjustment code}
\begin{sloppypar}
The program used for the orbit adjustment, ORBIT, derives from the
code of Tokovinin (1992). Like its progenitor, it performs a least
square adjustment to all available spectroscopic and ``visual''
observations, with weights inversely proportional to the square of
their standard errors. Besides a number of cosmetic changes, the
modifications and additions to Tokovinin (1992) include allowance for
multiple radial velocity zero-points, the use of the more robust
Levenberg-Marquardt minimisation algorithm, as well as direct support
for triple systems. Standard errors for derived parameters (masses,
parallax, etc) are computed from the full covariance matrix of the
linearized least square adjustment, rather than from just the standard
errors of the orbital elements (i.e. the diagonal terms of the
covariance matrix).  In addition ORBIT can estimate confidence
intervals for both derived parameters and orbital elements through
Mon{\-}te-Car{\-}lo experiments. For well constrained binaries (like Gl~570BC)
these intervals are consistent with the analytic standard errors and
gaussian statistics, incidentally providing a check against gross
errors in the analytic gradients. For noisier systems though, the
vicinity of the minimum of the ${\chi}^2$ surface becomes
significantly non-quadratic.  The error statistics for the parameters
are then substantially non-gaussian and asymetric, and the Monte-Carlo 
confidence intervals become an essential feature.
\end{sloppypar}

\begin{sloppypar}
ORBIT also accepts some additional data types, beyond spectroscopic
velocities and ``visual'' $({\rho},{\theta})$ or (X,Y) pairs. These additions,
all of which are used in the present paper, include:
\begin{itemize}
\item{} 1D projected separation measurements, produced by early IR
scanning speckle observations and lunar occultations;
\item{} parallaxes, which can be seen as an additional link between
the velocity amplitudes, the eccentricity, the period and the
semi-major axis; they enter in the least square adjustment with their
standard error, providing an optimal unified description between orbital 
parallaxes of SB2+visual pairs and mass ratios from SB1+visual pairs
with a known parallax;
\item{} cross-correlation ``dips'', produced by correlation
velocimeters and cross-correlation analysis of echelle spectra; to
support these profiles ORBIT complements the usual list of 10 orbital
elements by the width and the equivalent width of one gaussian profile
for each spectroscopic component, and adjusts the 10 orbital elements
and the profile parameters directly to the ensemble of all
cross-correlations, rho-theta pairs, parallaxes, etc. In a similar
manner, modern analyses of long baseline interferometric data
sometimes adjust orbital elements (together with magnitude
differences) directly to the {\it uv} data, avoiding the intermediate step
of $\rho, \theta$ extraction (Hummel \& Amstrong 1992), as do
adjustments of visual orbits to Hipparcos Transit Data (Quist \&
Lindegren 1999; S\"oderhjelm 1999). A practical inconvenient of this
bypassing of intermediate steps is that no individual radial
velocities (or $\rho, \theta$, or separations on great circles) are
available for compact publication, and that the usual figures of
radial velocity curve (or visual orbit) loose sense and become
illustrative only. On the positive side however, direct orbital
adjustment to the profiles greatly decreases the effective number of
degrees of freedom, compared with orbital adjustment to velocities
measured from individual profiles. We found that this often improves
the error bars on the spectroscopic orbital elements by a factor of
two, or better. Global adjustment also considerably reduces the
susceptibility of the orbital solution to ``pulling'' of the velocity
of the weaker component towards and away from that of its brighter
companion for profiles which are only imperfectly gaussian. In
addition, and to further diminish this sensitivity, we produce an
average recentered profile for each component, determine residuals to
the best fitting gaussian, and subtract those residuals from the
individual measurements to then determine iteratively improved
parameters.
\end{itemize}
ORBIT runs on Unix systems and can be obtained from the authors upon
request.  
\end{sloppypar}

\begin{sloppypar}
To illustrate our methods and the power of combining different data
types for the same binary, as probably first advocated by Morbey
(1975), we present 3 distinct orbits for Gl~570BC, incorporating
successively more of the information available to us. The speckle and
adaptive optics data alone turn out to be insufficient to properly
define a visual orbit, largely because they never resolve this
eccentric system close to its periastron (Fig.~2), where the minimal
separation is only $\sim$12 milliarcsecond. As a consequence these
data alone leave indeterminate a combination of inclination and
semi-major axis with eccentricity. The most restrictive orbit we can
present in Table~\ref{tab_orbit} is therefore a spectroscopic one. We
then present a combined spectroscopic and visual orbit, and finally an
orbit which in addition uses the independent trigonometric parallax of
the Gl~570 system.
\end{sloppypar}

The longitude of the periastron is given with the spectroscopic
convention, and thus refers to the primary. 180\deg must be added to
$\omega$ to obtain the visual convention. Table~\ref{tab_oa}
lists the individual speckle and adaptive optics measurements 
and Table~2  
(only available electronically) gives the same
information for the radial velocities. Note, though, that the final
orbit was directly adjusted to the individual cross-correlation
profiles, while the velocities given in Table~2 were determined
separately by adjusting two Gaussian curves to each profile. These
velocities are given here mostly for illustrative purposes, as plotted
in Fig. 1. Their use would give a slightly different (and noisier)
solution, but may nonetheless prove convenient. There is no compact
way to publish the full information needed to reproduce our analysis,
since it does not proceed through determining radial velocities as an
intermediate step. We will make the digital correlation profiles
available upon request to the first author. Fig.~2 shows the visual
data and orbit.

\subsection{Spectroscopic orbit}
As mentioned above, the spectroscopic orbit presented in
Table~\ref{tab_orbit} was adjusted directly to the correlation
profiles rather than to the extracted radial velocities.  As can be
seen, the spectroscopic quantities $M{\times}sin^{3}i$ are determined
with very high accuracies of 0.2\%. The relatively few ELODIE
measurements contribute considerably to the overall precision of the
orbital solution, and ignoring them would for instance degrade the
standard errors of the $M{\times}sin^{3}i$ by an order of
magnitude. The much more numerous CORAVEL measurements of the primary
star by contrast only contribute to an improved orbital period, thanks
to their much longer timespan. They otherwise carry very little weight
in the solution.

\begin{figure}
\psfig{width=8.5cm,file=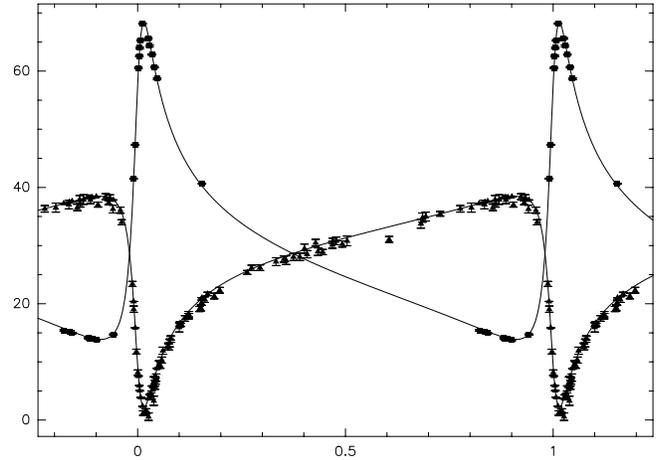,angle=-90}
\caption{Radial velocity orbit of the Gl~570BC system. Filled triangles
represent measurements of the primary star, and filled squares measurements
of the secundary stars. The ELODIE mesurements can be distinguished from the
CORAVEL ones by their much smaller error bars.
         }
\label{RV_orbit}%
\end{figure}

\begin{table}
  \center
  \tabcolsep 0.1cm
  \caption{Orbital elements and derived parameters}
  \begin{tabular}{lccc}
Element & Spectro & Spectro+OA & Spectro+OA+$\pi$\\[5pt]
V$_0$ (km/s) & 28.664$\pm$0.008  &28.665$\pm$0.008 &28.665$\pm$0.008 \\
P (days)     & 308.884$\pm$0.004   &308.884$\pm$0.004 &308.884$\pm$0.004 \\
T$_0$ (JD)   & 270.220$\pm$0.011 & 270.220$\pm$0.011& 270.220$\pm$0.011 \\
{\it e}      & 0.7559$\pm$0.0002 & 0.7559$\pm$0.0002& 0.7559$\pm$0.0002\\
{\it a} (arcsec) & -- &0.1514$\pm$0.0015 &0.1507$\pm$0.0007  \\
${\Omega}_1$ (deg) & -- & 196.2$\pm$0.7&195.9$\pm$0.5 \\
${\omega}_1$ (deg) & 127.56$\pm$0.05 &127.56$\pm$0.05 &127.56$\pm$0.05\\
{\it i} (deg)  & -- & 107.2$\pm$1.0 & 107.6$\pm$0.7 \\
K$_1$ (km/s) & 18.187$\pm$0.008 &18.187$\pm$0.008 & 18.187$\pm$0.008\\
K$_2$ (km/s) & 27.325$\pm$0.026 &27.324$\pm$0.026 & 27.325$\pm$0.026\\
W$_1$ (km/s) &  6.169$\pm$0.008 & 6.169$\pm$0.008 & 6.169$\pm$0.008\\
W$_2$ (km/s) &  5.643$\pm$0.030 &  5.642$\pm$0.030& 5.642$\pm$0.030  \\
EW$_1$ (km/s) & 0.4532$\pm$0.0005 &0.4532$\pm$0.0005 & 0.4532$\pm$0.0005\\
EW$_2$ (km/s) & 0.1041$\pm$0.0005& 0.1041$\pm$0.0005 & 0.1041$\pm$0.0005\\
${\Delta}$V$_{CRV-ELO}$ &-0.633$\pm$0.065 & -0.633$\pm$0.065 & -0.633$\pm$0.065\\
 \\
Derived parameters: \\
M$_1{\times}sin^{3}i (\Msol) $ & 0.5082$\pm$0.0011 & 0.5082$\pm$0.0011 &
0.5083$\pm$0.0011\\
M$_2{\times}sin^{3}i (\Msol) $ & 0.3383$\pm$0.0005 & 0.3383$\pm$0.0005 &
0.3383$\pm$0.0005\\
M$_1$ (\Msol) &  -- &0.583$\pm$0.009  & 0.586$\pm$0.007 \\
M$_2$ (\Msol) &  -- &0.388$\pm$0.006 & 0.390$\pm$0.005\\
$\pi$ (arcsec) & -- & 0.1710$\pm$0.0022 & 0.1698$\pm$0.0009 \\
  \end{tabular}
\raggedright
\noindent
\vskip 2mm
Notes: Epochs are listed as offsets relative to Julian Day
2450000. W$_1$ and W$_2$ are the full widths to half power of the
Gaussian function for the two components, while EW$_1$ and and EW$_2$
are their equivalent widths. The other orbital elements have their
usual meaning.
\label{tab_orbit}
\end{table}

\subsection{Spectroscopic+visual orbit}
Inclusion of the speckle and adaptive optics data in the adjustment
leaves all spectroscopic elements essentially unchanged
(Table~\ref{tab_orbit}), but determines the three otherwise unknown
orbital elements: the semi-major axis ({\it a}), the inclination ({\it i}),
and the orientation of the projected orbit on the sky
($\Omega$). Individual stellar masses as well as an orbital parallax
for the system can then be derived from the full complement of orbital
elements, and are also listed in Table~\ref{tab_orbit}. The standard
errors computed from the covariance matrix
were checked through Monte-Carlo simulations, which resulted in
confidence intervals that were fully consistent with gaussian errors
of the stated dispersion. This verifies that non-linearities in the
least square adjustment are negligible for this very well constrained
system. The two masses are determined here with 1.6\% accuracy, and the 
orbital parallax to within 2.2~milliarsecond (1.3\%).

\begin{sloppypar}
The availability of Hipparcos parallaxes represents an opportunity to
independently verify the orbital parallax, and thus to globally check
the orbital solution for systematic errors. Though somewhat noisier than 
typical for a V=5.7 star, the Hipparcos catalog parallax of Gl~570A is well
determined,
$\pi$=0.1693$\pm$0.0018$''$.  The parallax for Gl~570BC itself has very 
large error bars (${\sigma}({\pi})$=0.033$''$) in the Hipparcos
catalog (ESA 1997), because the unaccounted orbital motion
with P$\sim$0.8~year strongly couples into the parallax solution over
the limited lifetime of the Hipparcos satellite. Fortunately,
S\"oderhjelm (1999) recently reanalysed the Hipparcos intermediate transit
data, accounting for the orbital motion within Gl~570BC, and obtained
sharply reduced error bars for the trigonometric parallax:
${\pi}$=0.1697$''{\pm}$0.0010$''$.  The two astrometric determinations are
mutually consistent and agree with the orbital
parallax of 0.1710$\pm$0.0022$''$, 
to within better than 1~$\sigma$.
\end{sloppypar}
\begin{figure}
\psfig{width=80mm,file=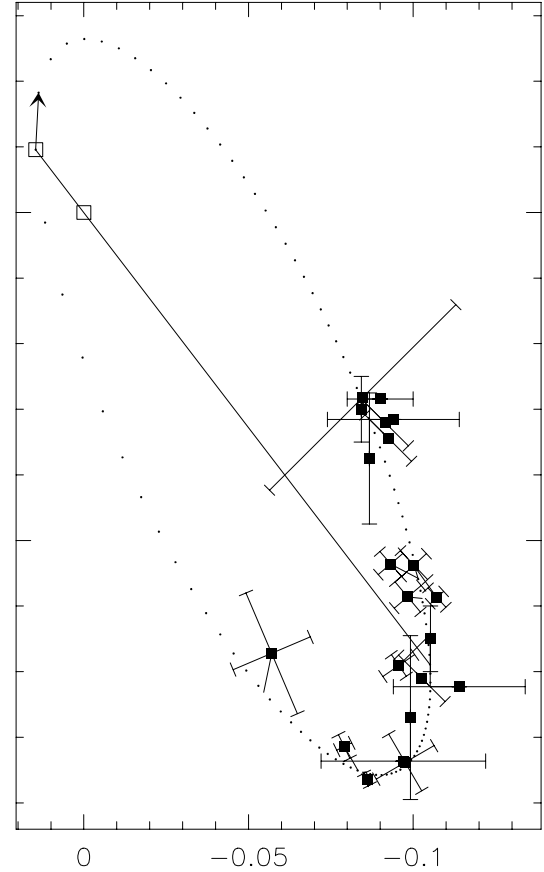,angle=-90}
\caption{Visual orbit of the Gl~570BC system. Separations are in arc 
seconds, North is up and East is left.
         }
\label{visual_orbit}%
\end{figure}

\subsection{Spectroscopic+Visual+Parallax orbit}
Instead of using the independent trigonometric parallaxes as a sanity
check for the orbital solution, ORBIT offers the option to consider it
as an additional observation, linking {\it i}, {\it e}, {\it P}, {\it a} and
K$_1$+K$_2$. It is then included in the combined least square
adjustment, together with the radial velocities and angular
separations. This very significantly improves the determinacy of the
least square system, and in particular reduces the standard errors of
the semi-major axis
by a factor of 2. We note here that it would be preferable, at least
conceptually, to adjust an orbit directly to the Hipparcos transit
data rather than to the Hipparcos parallax. ORBIT does not yet support
this data type however, which we plan to add in the near future.

\begin{sloppypar}
The contributions to the overall $\chi^2$ of the
different data types included in the solution are approximately
consistent with their respective number of measurements. This
indicates that there are no large systematic errors in any one
data type, and that the adopted standard errors are at least
approximately correct. This orbit (last column of
Table~\ref{tab_orbit}) has the smallest errorbars (and smallest
covariances) for all orbital elements. It is consistent with all
previously published orbits to within their stated error bars, after
allowing for the 180\deg ambiguity in the identification of the
ascending node (i.e. $\Omega$) in purely visual orbits. We adopt it as
our preferred solution for the rest of the discussion. The inclusion
of the trigonometric parallax information improves the relative
accuracy of the two masses to 1.2\%:
M$_1$~=~0.586$\pm$0.007{\Msol} and M$_2$~=~0.390$\pm$0.005{\Msol}.
They are consistent with the values obtained by Mariotti et al. (1990)
to within a fraction of their quoted standard deviation, as well as
with the mass sum derived by S\"oderhjelm (1999) from an analysis of
the Hipparcos transit data,
but improve on their accuracy by over a factor of 5.
These accuracies are among the best obtained to date for non-eclipsing
binaries, but still do not match the ${\sim}0.2\%$ obtained for the
spectroscopic $M{\times}sin^{3}i$.  There is thus still room for
improvement in the astrometric measurements. We are conducting some
long baseline interferometric observations of Gl~570BC (Segransan et
al. in progress) with IOTA (Millan-Gabet et al. 1999), which have the
potential to ultimately match the sub-\% spectroscopic accuracy for
the masses. The ``resulting'' parallax in the last column of
Table~\ref{tab_orbit}), 0.1698$\pm$0.0009$''$, is computed as an
orbital parallax from the elements of this combined orbit, which
themselves take the trigonometric parallax into account. It is an
optimal combination of the distance information available in the
trigonometric parallax and in the orbit. In the present case is
mostly determined by the weighted average of the two HIPPARCOS
parallaxes, with little contribution from the orbit.
\end{sloppypar}

\section{Luminosities and colours}
Photometric measurements of the combined light of the system have been made
by a number of authors over a broad range of colours. They are summarised
and homogenized in the critical compilation of Leggett (1992), from which
we adopt: U=10.79, B=9.57 ,V=8.09, R=7.09, I=5.97, J=4.75, H=4.14, 
K=3.93, \linebreak  L=3.77, L'=3.67, with quoted errors of $<$3\% on IJHK.
The system has colours consistent with those of solar metallicity
stars (Leggett 1992), as expected from the spectroscopic metallicity
discussed below.

\begin{table}
  \center
  \tabcolsep 0.1cm
  \caption{Absolute magnitudes of the two components}
  \begin{tabular}{|l|c|c|}
    \hline
    & Primary & Secondary \\
    \hline
    M$_{{J}}$ & 6.213$\pm$0.033  & 7.403 $\pm$0.039 \\
    M$_{{H}}$ & 5.613$\pm$0.033  & 6.763 $\pm$0.039 \\
    M$_{{K}}$ & 5.396$\pm$0.033  & 6.576 $\pm$0.039 \\
    M$_{{L}}$ & 5.167$\pm$0.111  & 6.227 $\pm$0.159 \\
    \hline
  \end{tabular}
  \label{tab_mag}
\end{table}

\begin{sloppypar}
The difference in brightness between the components of Gl~570BC has
been measured on several occasions in different bandpasses.  Mariotti
et al. (1990) obtained magnitude differences of
${\Delta}_J$=1.27$\pm$0.12, ${\Delta}_H$=1.19$\pm$0.12,
${\Delta}_K$=1.12$\pm$0.07, and ${\Delta}_L'$=1.06$\pm$0.17 from their
1D scanning speckle observations.  HMcC published another estimate in
the J band, ${\Delta}_J$=1.30$\pm$0.04. Our own adaptive optics
measurements provide ${\Delta}_K$=1.18$\pm$0.04, as well as 
${\Delta}_m$=1.18$\pm$0.04 in a narrow band Br$_{\gamma}$ 
(2.166${\mu}m$) filter. We estimate a correction of 
${\Delta}_K-{\Delta}_{Br_{\gamma}}$=0.04, using synthetic photometry generated
from NextGen (Hauschild, Allard, \& Baron 1999) model spectra
of effective temperatures 
that bracket those apropriate for M dwarfs of the luminosities of the
two components of Gl~570BC. This provides a second 
determination of the K band flux ratio, ${\Delta}_K$=1.22$\pm$0.04. 
\end{sloppypar} 

\begin{sloppypar}
While those measurements are mutually consistent for every filter,
their run with wavelenghth is only marginally compatible with the
known IR colours of M dwarfs (Leggett 1992). The $J-H$ and
(particularly) $J-K$ colours of early and mid-M dwarfs define a
remarkably flat plateau (Leggett 1992): when the spectral type of solar 
metallicity stars (apropriate for Gl~570BC) varies between M0V
(M$_K\sim$4.8) and M5.5V (M$_K\sim$7.9), $J-H$ only changes from 0.68 to 
0.57, and $J-K$ just from 0.85 to 0.87 (Leggett 1992)
The two components of Gl~570BC respectively have M$_K\sim$5.4 and
M$_K\sim$6.6 and their $J-H$ and $J-K$ colours are therefore expected to
only differ by about -0.04 and +0.01. ${\Delta}_J$ should therefore be
almost identical to ${\Delta}_K$, whereas the observations give
${\Delta}_J$-${\Delta}_K$=0.14$\pm$0.04, and ${\Delta}_J$-${\Delta}_H$
should be -0.04, whereas the observations indicate
${\Delta}_J$-${\Delta}_H$=+0.11$\pm$0.12.
As the integrated colours of the system match the expected values,
this inconsistency must come from some of the measured magnitude
differences.  We suspect that it traces back to an overestimated
contrast in the J and H band observations of both Mariotti et
al. (1990) and HMcC, since speckle techniques have a known bias
towards underestimating the relative flux of faint components (Perrier
1988). The K band adaptive optics observations are in principle immune
to this bias, and the K band speckle observations of Mariotti et
al. (1990) should be relatively unaffected, thanks to the low D/$r_0$
(where D is the telescope diameter and $r_0$ is the Fried parameter of
the atmosphere) at this longer wavelength. We therefore tentatively
adopt as the basis of our magnitude difference system the mean of the
three K band measurements, ${\Delta}_K$=1.18$\pm$0.03. From this value
we then derive preferred values of ${\Delta}_J$=1.19 and
${\Delta}_H$=1.15, but we will also consider the published J and K flux
ratios as an alternative.
This discrepancy contributes significant uncertainty to the analysis,
and better measurements of the flux ratios at $J$ and $H$ would be of
considerable interest. To date all our adaptive optics measurements in
the $J$ and $H$ bands have unfortunately been obtained at phases when the
secondary star overlaps the first Airy ring of the primary for these
wavelengths. These circumstances maximize the uncertainties in
differential photometry from partially corrected adaptive optics
images (Veran et al. 1999), and these data therefore contribute no
useful information on the flux density ratio.
\end{sloppypar}

Absolute magnitudes were derived from the parallax and the apparent
magnitudes of the individual stars.  The absolute magnitudes
of the brighter star have uncertainties which are dominated by those
of the photometry of the system, while those for the secondary have
some contribution from the magnitude difference. The parallax doesn't
appreciably contribute in either case. 

\section{Discussion and conclusions}
The accurate masses and absolute magnitudes that we have obtained for
the Gl~570BC system represent a new benchmark for model calculations
(e.g. Baraffe et al.  1998) and an independent check of the empirical
mass-luminosity relations (HMcC). The constraints which they bring to
the models are largely complementary to those coming from the eclipsing
binaries, whose absolute radii can be determined very accurately but
whose larger distances on the other hand contribute significant 
uncertainties to the absolute magnitudes.

As long emphasized by theoreticians, and by observers of more massive
stars (e.g. Andersen 1991), there is however no such thing as one
single mass-luminosity relation: stellar luminosities depend on
chemical composition as well as on mass (in general they depend on age
too, though not in the age and mass range discussed here).
Quantitative metallicity determinations however are notoriously
difficult for M dwarfs (e.g. Viti et al. 1997; Valenti et
al. 1998). Observers in this field usually have to resort to
photometric metallicity estimators (Leggett 1992) which are only
approximately calibrated, or otherwise assume by default a solar
metallicity.  Thanks to its physical association with the hotter
Gl~570A (K4V), the Gl~570BC pair represents a rare case of
two M dwarfs with a very well determined spectroscopic
metallicity. Its accurate masses are thus fortunately matched with
excellent metallicities.  Hearnshaw (1976) first measured the
metallicity of Gl~570A from high resolution electronographic spectra
and obtained [Fe/H]=+0.01$\pm$0.15.
More recently, Feltzing \& Gustafsson (1998) measured \linebreak[4] 
[Fe/H]=0.04$\pm$0.02(random)$\pm$0.1(systematic) from high S/N \linebreak[4] 
R=10$^5$ echelle CCD spectra. These authors find some evidence for
NLTE overionisation of Fe into Fe$^+$, but the derived elemental Fe
abundance is unaffected, as Fe is overwhelmingly neutral in the
photosphere of a K4 dwarf. Quite conveniently for comparison with published
models, the Gl~570 system thus has a truly solar metallicity.

%

\begin{figure}
  \psfig{width=85mm,file=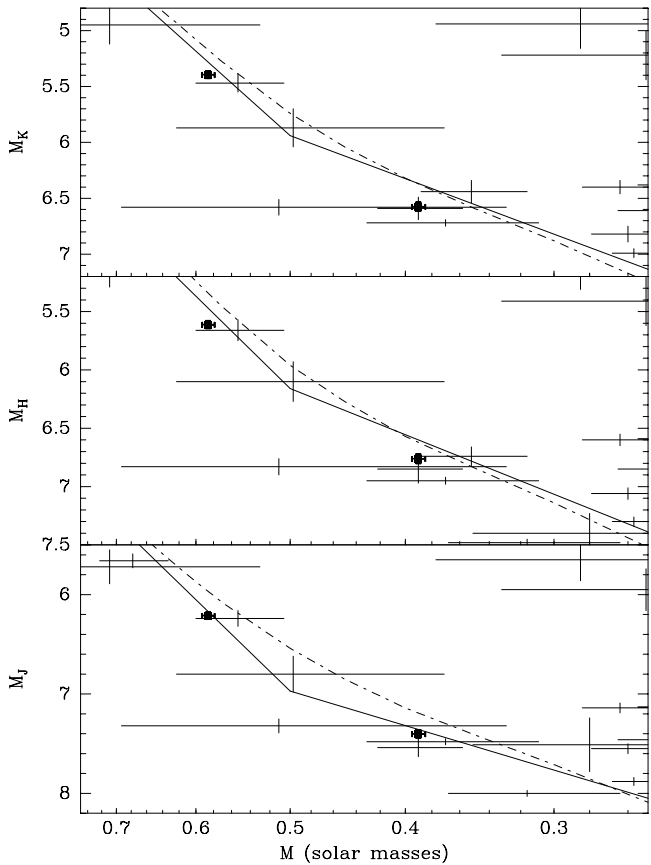,angle=0}
  \caption{Near-IR Mass-Luminosity diagrams for mid-M dwarfs. The square
    symbols with thick error bars correspond to the two components of
    Gl~570BC, and the thin error-bars represent data from HMcC. The solid
    and dot-dashed lines respectively represent the empirical HMcC 
    analytical Mass-Luminosity relation and the theoretical [M/H]=0.0 10~Gyr
    isochrone of Baraffe et al. (1998).
    }%
  \label{ML_plot}%
\end{figure}
%

\begin{sloppypar}
Figure \ref{ML_plot} compares the mass and luminosity of the two
components of Gl~570BC with the Baraffe et al. (1998) 10~Gyr
solar-metallicity isochrone. These models consistently combine stellar
evolution models (e.g. Chabrier \& Baraffe 1997) and non-grey
atmospheric models (Allard \& Hauschild 1995; Hauschild et al. 1999).
The present generation of these evolutionary models still uses
non-dusty atmospheres, but dust only becomes relevant at effective
temperatures significantly lower than those of Gl~570BC (Allard
1998). The Baraffe et al. models are consistently slightly brighter in
all 3 bands than the two stars, by 0.08 to 0.15 magnitude.  While this
level of agreement is already very comforting, the discrepancies are
significant at the $\sim$3$\sigma$ levels and may point towards
remaining low level deficiencies of the theoretical models. If one
adopts the measured J and H band flux ratios, rather than the smaller
values that we have estimated from the K band measurement, the
agreement with the models improves only slightly for the primary star,
as its magnitude only weakly depends on the exact value of the large
flux ratio. At the same time this choice degrades the agreement for the
fainter star by a larger factor, and the overall agreement with the
models is significantly worse.
\end{sloppypar}


Figure \ref{ML_plot} also shows the data points of HMcC, as well as
their analytic representation of those data. The agreement is
essentially perfect with the $J$ band HMcC relation, while the $H$ and 
$K$ band relations are slightly discrepant, by respectively 0.1 and 0.15
magnitudes.  We note that the HMcC mass-luminosity relations are only
consistent with the empirical M dwarf colours of Leggett
(1992) at this 0.10--0.15~magnitude level, even though the HMcC
photometry mostly traces back to Leggett (1992). This is because the
HMcC mass-magnitude relations for $J$, $H$ and $K$ were adjusted
independently, without explicit forcing of colour consistency. The
perfect agreement with the $J$ band HMcC fit is therefore probably
fortuitous to some extent, and the 0.10 to 0.15 discrepancy for the $H$
and $K$ bands probably represents a more realistic estimate of the
accuracy of those analytic fits around 0.5~{\Msol}.

\begin{acknowledgements}
  We dedicate this paper to the memory of our two late colleagues,
  Antoine Duquennoy and Jean Marie Mariotti, who started this program
  with two of us (CP, MM) and pushed it forward until their untimely death. 

\begin{sloppypar}
  We are indebted to the many CORAVEL observers who contributed to
  measurements of Gl~570BC over the years, and to Didier Queloz,
  Dominique Naef and Nuno Santos who obtained some ELODIE and CORALIE
  radial velocities for us at critical orbital phases.
  We also warmly thank the colleagues who assisted us for the 2D
  speckle or AO observations : Steve Ridgway and Julian Christou who
  participated in the KPNO 2D speckle observation and Ron Probst for
  the operation of the infrared imager, Daniel Rouan, Fran\c{c}ois
  Lacombe and their group for the assistance with the CFHT CIRCUS
  infrared camera and E. Tessier who wrote and operated its speckle
  acquisition and reduction software, the operators and the support and
  infrared teams of ESO, La Silla for their help during COME-ON+
  observations. We are grateful to the anonymous referee for
  constructive comments and clarification requests, which led to an
  improved paper.
\end{sloppypar}
\end{acknowledgements}

\newpage
\hbox{}
\newpage

\setcounter{table}{1}
\begin{tabular}{|l|ll|} \hline
Julian Day   &   V$_1$               &     V$_2$   \\ 
2400000+ & (in km/s) & (in km/s) \\ \hline \hline
\multicolumn{3}{c}{CORAVEL measurement} \\ \hline
43578.641 &  27.250 $\pm$   .680 & \\
43591.613 &  28.040 $\pm$   .610  & \\
43631.521 &  30.800 $\pm$   .830 & \\
43686.390 &  33.750 $\pm$   .750  & \\
43908.694 &  29.470 $\pm$   .790 & \\
43996.501 &  34.810 $\pm$   .780  & \\
44024.440 &  36.310 $\pm$   .570 & \\
44042.420 &  37.070 $\pm$   .760 & \\
44053.380 &  37.910 $\pm$   .830 & \\
44307.656 &  34.990 $\pm$   .770 & \\
44341.571 &  36.520 $\pm$   .780   & \\
43918.672 &  28.960 $\pm$   .650 & \\
44690.552 &  38.240 $\pm$   .510 & \\
45001.730 &  36.270 $\pm$   .660  & \\
45044.641 &  13.930 $\pm$   .560   & \\
45130.422 &  27.490 $\pm$   .730 & \\
45152.383 &  30.470 $\pm$   .590 & \\
45413.611 &  26.180 $\pm$   .460 & \\
45437.585 &  27.680 $\pm$   .520 & \\
45449.524 &  28.050 $\pm$   .620    & \\
45466.479 &  28.900 $\pm$   .450 & \\
45476.530 &  30.800 $\pm$   .670 & \\
45728.740 &  26.210 $\pm$   .470 & \\
45763.694 &  28.520 $\pm$   .620  & \\
45782.615 &  30.130 $\pm$   .590 & \\
46133.650 &  31.070 $\pm$   .500  & \\
46212.426 &  37.910 $\pm$   .700  & \\
46220.405 &  37.840 $\pm$   .630 & \\
46267.353 &   6.500 $\pm$   .920 & \\
46268.346 &   6.580 $\pm$   .770 & \\
46268.356 &   6.050 $\pm$   .770 & \\
46271.357 &   9.630 $\pm$   .570  & \\
46272.355 &   9.400 $\pm$   .520  & \\
46274.347 &  11.930 $\pm$   .610  & \\
46278.334 &  12.640 $\pm$   .620 & \\
46279.342 &  13.270 $\pm$   .680 & \\
46269.345 &   7.290 $\pm$   .630 & \\
46519.571 &  36.560 $\pm$   .540 & \\
46521.618 &  37.110 $\pm$   .500  & \\
46540.557 &  38.410 $\pm$   .510 & \\
46551.513 &  36.080 $\pm$   .530  & \\
46552.492 &  34.020 $\pm$   .450  & \\
46560.473 &  23.410 $\pm$   .440 & \\
46561.475 &  19.070 $\pm$   .540 & \\
46563.485 &  11.780 $\pm$   .410 & \\
46564.489 &   8.180 $\pm$   .450  & \\
46568.454 &   1.200 $\pm$   .380  & \\
46569.457 &   1.170 $\pm$   .430 & \\ \hline
\end{tabular}

\begin{tabular}{|l|ll|} \hline
Julian Day   &   V$_1$               &     V$_2$   \\ 
2400000+ & (in km/s) & (in km/s) \\ \hline \hline
\multicolumn{3}{c}{CORAVEL measurement (continuation)} \\ \hline
46570.436 &   1.460 $\pm$   .460 & \\
46572.456 &    .580 $\pm$   .640  & \\
46574.456 &   4.100 $\pm$   .590 & \\
46576.449 &   3.310 $\pm$   .780 & \\
46582.437 &  10.310 $\pm$   .520 & \\
46587.414 &  12.590 $\pm$   .520 & \\
46595.357 &  15.960 $\pm$   .810 & \\
46602.395 &  18.200 $\pm$   .470 & \\
46602.423 &  17.860 $\pm$   .470  & \\
46611.413 &  19.290 $\pm$   .520  & \\
46621.389 &  21.320 $\pm$   .490 & \\
46883.575 &   4.330 $\pm$   .680   & \\
46886.604 &   6.730 $\pm$   .570  & \\
46908.527 &  17.470 $\pm$   .480 & \\
46305.497 &  20.700 $\pm$   .480 & \\
46316.488 &  22.340 $\pm$   .490 & \\
48024.672 &  35.460 $\pm$   .410 & \\
48448.664 &  16.380 $\pm$   .310 & \\
48449.603 &  16.360 $\pm$   .290 & \\
48463.505 &  19.090 $\pm$   .450 & \\
48469.573 &  21.680 $\pm$   .310 & \\
48696.881 &  37.040 $\pm$   .430 & \\
48704.848 &  37.380 $\pm$   .330  & \\
48732.747 &   1.630  $\pm$  .320  & \\
48871.507 &  30.790 $\pm$   .470  & \\
48878.514 &  30.270 $\pm$   .540 & \\
49059.893 &  14.070 $\pm$   .310 & \\
49066.821 &  16.650 $\pm$   .300 & \\
49082.824 &  20.190 $\pm$   .450 & \\
49116.709 &  25.390  $\pm$  .340  & \\ \hline
\multicolumn{3}{c}{ELODIE measurement} \\ \hline
50524.647 &  37.252 $\pm$   .029 &  15.423  $\pm$  .124   \\
50561.530 &  37.843  $\pm$  .026 &  14.718  $\pm$  .099   \\  
50576.491 &  20.394 $\pm$   .027 &  41.514  $\pm$  .093   \\  
50587.458 &   3.679 $\pm$   .034 &  65.622  $\pm$  .148   \\  
50588.455 &   4.366 $\pm$   .038 &  64.358  $\pm$  .156   \\   
50627.353 &  21.035 $\pm$   .028 &  40.644  $\pm$  .097    \\ 
50837.711 &  37.502 $\pm$   .031 &  15.162  $\pm$  .144    \\ 
50839.715 &  37.627 $\pm$   .033 &  14.939  $\pm$  .140    \\ 
50851.704 &  38.237 $\pm$   .031 &  14.096  $\pm$  .117   \\  
50852.707 &  38.283 $\pm$   .035 &  14.000  $\pm$  .126   \\   
50853.704 &  38.273 $\pm$   .033 &  13.919  $\pm$  .127   \\  
50854.704 &  38.309 $\pm$   .030 &  13.906  $\pm$  .121   \\   
50857.713 &  38.417 $\pm$   .033 &  13.769  $\pm$  .129   \\   
50886.622 &  15.867 $\pm$   .023 &  47.349  $\pm$  .086    \\ 
50889.637 &   5.941 $\pm$   .036 &  62.600  $\pm$  .125    \\  
50890.645 &   3.907 $\pm$   .033 &  65.285  $\pm$  .131   \\   \hline
\end{tabular}

\begin{tabular}{|l|ll|} \hline
Julian Day   &   V$_1$               &     V$_2$   \\ 
2400000+ & (in km/s) & (in km/s) \\ \hline \hline
\multicolumn{3}{c}{CORALIE measurement} \\ \hline
51197.874 &   7.546  $\pm$  .030 &  60.549  $\pm$  .120   \\
51198.870 &   5.057 $\pm$   .030 &  63.986  $\pm$  .130    \\ 
51200.880 &   2.391 $\pm$   .030 &  68.186  $\pm$  .120   \\
51207.859 &   5.868 $\pm$   .030 &  62.853  $\pm$  .120   \\
51209.886 &   7.488 $\pm$   .030 &  60.661  $\pm$  .130   \\
51211.778 &   8.883 $\pm$   .030 &  58.694  $\pm$  .120   \\ \hline
\end{tabular}

\end{document}